\documentclass[10pt]{article}

\usepackage{graphicx}

\def\b{\begin{eqnarray}}

\def\e{\end{eqnarray}}

\usepackage{geometry}
\begin{document}


\begin{center}
{\LARGE \bf Hawking radiation screening and
\vspace{5mm}

 Penrose process shielding in the Kerr

\vspace{5mm}
Black Hole}

\end{center}

\vspace{5mm}

\begin{center}

{ \large\bf Eamon Mc Caughey}

\end{center}

\begin{center}
{\small \textit { School of Mathematical Sciences }}

{\small \textit{ Dublin Institute of Technology, Dublin 8, Ireland }}

{\small \textit { email: eamon.mccaughey@dit.ie }}

\end{center}

\begin{center}

{\large\bf Abstract }

\end{center}

\begin{center}
\begin{minipage}{0.85\textwidth}

{\footnotesize The radial motion of  massive  particles in the equatorial plane of a Kerr black hole is considered. Screening of the Hawking radiation and shielding of the Penrose process are examined (both inside and outside the ergosphere) and  their effect on the evaporation of the black hole is studied. In particular, the locus and width of a classically forbidden region and their dependance on the particle's angular momentum and energy is analysed. Tunneling of particles between the boundaries of this region is considered and the transmission coefficient determined.

keywords: Kerr geometry; black hole decay; quantum tunnelling. }

\end{minipage}
\end{center}

\section{Introduction }

 Dimopoulos and Landsberg \cite{Dimopoulos1}, and Giddings and Thomas \cite{Giddings1} suggested that, as a consequence of TeV-scale quantum gravity, the production of microscopic black holes at the LHC  might be possible. These black holes would decay rapidly into charged leptons and photons with a clean signature and low background.
Microscopic black holes offer the possibility of gaining insights into quantum gravity and contribute  to efforts to reconcile general relativity and quantum theory. Once the energy threshold is crossed (Planck mass $\approx$  1 TeV ),  the production of black holes is possible. However, at this scale unknown quantum effects play an important role and the black holes become "stringy"  and complex.
In the present analysis quantum gravity effects are ignored in favour of classical arguments.
   Black holes decay in four distinct phases: the balding phase, the spin down phase, the Schwarzschild phase and the
Planck phase. In this work, the spin-down phase is only considered---this is the phase in which case the Kerr black hole evolves into a Schwarzschild black hole by losing mass through Hawking radiation \cite{Hawking1,Hawking2} and angular momentum through the Penrose process\cite{Penrose}. Radial motion is characterised by a distinct region with negative kinetic energy for particles with certain angular momentum and energy. Classically, a particle cannot be in such a region, thus, the boundaries of this forbidden region are turning points. This paper studies the effects of the negative kinetic energy region on the spin-down phase of a Kerr black hole.

Decay processes in macroscopic black holes  were first studied by Hawking and led to the prediction of Hawking radiation. Parikh and Wilczek \cite{Parikh1} developed this idea as a quantum tunneling process, across a classically forbidden region at the horizon. Prodanov \cite {Prodanov1} has recently examined the screening of such evaporation in the Schwarschild and Reissner-Nordstrom cases.

  The reduction of the mass of a black hole is, in part, due to Hawking radiation. Emitted particles with certain ranges of energies and angular momenta may be reflected back into the black hole by a turning point outside the event horizon, thus screening of the evaporation takes place. The screening, however, is offset by a competing process: quantum tunneling across the forbidden region. This tunneling reduces the screening effect on the evaporation. The overall effect is the damping of a range of Hawking radiation spectral modes.
For particles with lower angular momentum, the screening effect is less pronounced and tunneling is more likely. Therefore, the Hawking radiation spectrum is dominated by low angular momentum modes and high energy modes.

The Penrose process is responsible for the loss of angular momentum of a rotating black hole. Particles entering the ergosphere may be scattered and ejected with greater energy. This process is shielded by the negative kinetic energy region for particles with certain ranges of their energies and  angular momenta. For particles with large angular momentum and low energy this region extends beyond the static limit and "swallows" the ergosphere. The black hole is completely shielded from such particles and the Penrose process is eliminated. If the negative energy region does not extend to the static limit only partial shielding takes place. In both cases particles can tunnel across  the forbidden region in the ergosphere and may or may not contribute to energy extraction. If particle splitting occurs between the event horizon and the lower boundary of the forbidden region,which is never between the horizons, energy extraction can only take place if the particle tunnels  back across the potential barrier. For higher angular momentum modes the potential barrier moves ever closer to the event horizon making this effect more and more unlikely.

Since Penrose proposed his process for energy extraction from a Kerr black hole many authors have suggested alternative mechanisms. The efficiency of the original process was estimated by Wald to be in the region of 120\% \cite{Wald}. Banados, Silk and West (BSW)\cite{Banados} have demonstrated that colliding particles near the event horizon of an extremal black hole can result in higher energy extraction efficiency due to the high centre-of-mass energy of the collision. However, several authors subsequently have shown that the efficiency of this process is limited to 130 \%. More recently, a new limit of energy extraction efficiency has been demonstrated by Schnittman. For Compton scattering between a photon and a massive particle, the efficiency can reach 1400\% \cite{Schnittman}. High centre-of-mass collisions can produce very energetic particles making Kerr black holes act as natural particles accelerators.

The current work concerns trajectories of particles inside and outside the ergosphere of a  rotating black hole. This involves a detailed study of radial motion of particles in the Kerr  metric, in particular, the effect of the angular momentum and energy of the particles on the shielding process is studied. Tunneling of such particles between the boundaries of the classically forbidden region is examined and an analytical expression for the transmission coefficient across the barrier is determined.

\section{ The Kerr Metric and the Equations of Motion}

 The Kerr metric in Boyer-Lindquist
coordinates is given by
\cite{Frolov1}:

\b
  ds^2 &=& \frac{\Delta -a^2 \sin^2\theta}{\rho^2}\, dt^2+\frac{4Mar\sin^2\theta}{\rho^2}\,dt\,d\phi-\frac{\rho^2}{\Delta}\,dr^2 - \rho^2 \,d\theta^2   \\
  & & - \big[\frac{2Ma^2r\sin^2\theta}{\rho^2}+(r^2+a^2)\big]\sin^2\theta\, d\phi^2,  \nonumber
\e
where  $ \rho^2 = r^2 + a^2\cos^2\theta $,\,\,\, $\Delta = r^2 -2Mr + a^2,$
with $ a$ --- the specific angular momentum of the black hole and
 $M$ --- the mass of the black hole.

The Lagrangian for a particle of mass $m$ and charge $q$ in the presence of a gravitational and electromagnetic field $A$ is given by \cite{Frolov1}:

\b L=\frac{1}{2}g_{ij}\frac{dx^i}{d\lambda}\frac{dx^j}{d\lambda}+\frac{q}{m}A_i\frac{dx^i}{d\lambda},
\e
where $\lambda$ is the proper time $\tau$ per unit mass $m$: $\lambda=\tau/m $ and $A$ --- the electromagnetic potential given by:

\b
A_idx^i= \frac{Qr}{\Sigma^2}(dt-a \sin^2\theta\, d\phi). \nonumber
\e
In Boyer-Lindquist coordinates the Lagranian becomes:
\b
\label{Lagrangian}
L=\frac{1}{2}g_{tt}\dot{t}^2+\frac{1}{2}g_{rr}\dot{r}^2+\frac{1}{2}g_{\theta\theta}\dot{\theta}^2+\frac{1}{2}g_{\phi\phi}\dot{\phi}^2+\frac{qQr}{m\Sigma^2}\,\dot{t}-\frac{aqQr}{m\Sigma^2}\sin^2\theta\,\dot{\phi}.
\e
\newline
The motion of the particle is completely determined by the geodesic equation \cite{Frolov1}:

\b
\frac{d^2 x^i}{d\tau^2}+ \Gamma^i_{jk}\frac{dx^j}{d\tau}\frac{dx^k}{d\tau}=\frac{q}{m}F^i_j\frac{dx^j}{d\tau},
\e
where $ F= dA $ is Maxwell's electromagnetic tensor and $ \Gamma^i_{jk} $ are the Christoffel symbols.
This results in four equations of motion for the test particle \cite{Carter}:
\newline

\b
 \rho^2\frac{dt}{d\lambda}&=& -a^2 E \sin^2\theta + aJ + \frac{r^2+a^2}{\Delta}[E(r^2+a^2)-Ja], \\
  \label{Carter2}
 \rho^2\frac{dr}{d\lambda} &=& \sqrt{[E(r^2+a^2)-Ja]^2-\Delta[m^2r^2+(J-aE)^2+K]}, \\
 \rho^2\frac{d\theta}{d\lambda} &=& \sqrt{K-\cos^2\theta[a^2(m^2-E^2)+\frac{1}{\sin^2\theta}J^2]},  \\
 \rho^2\frac{d\phi}{d\lambda} &=& - aE+\frac{J}{\sin^2\theta}+\frac{a}{\Delta}[E(r^2+a^2)-Ja],
\e
where  $K$ is a conserved quantity (called Carter's constant) given by:

\b
K &=& p^2_\theta+ cos^2 \theta[a^2(m^2-E^2 )+ \frac{1}{\sin^2 \theta }J^2 ].
\e
\newline
In the above equation, $ E= (1/m)\partial L/\partial \dot{t}  $ is the conserved energy of the particle,
$ J= (1/m)\partial L/\partial \dot{\phi} $ is the conserved projection of the particle's angular momentum on the axis of rotation,
and
$ p_\theta =(1/m)\partial L/\partial \dot{\theta} $   is the $ \theta$-component of the particle's four-momentum.

For the present analysis, only radial motion in the equatorial plane is considered.  Therefore  $ \theta = \pi/2$ and $\dot{\theta} = 0$.
Hence $ K = 0 $.
Equation (\ref{Carter2}) gives the kinetic energy per unit mass of the particle:
\b
\label{KineticE}
\frac{\dot{r^2}}{2}=\frac{1}{2}(\epsilon^2 -1 )+\frac{1}{r}M+\frac{1}{2r^2}\big[a^2(\epsilon^2-1) -j^2\big]+ \frac{1}{r^3}M(a\epsilon-j)^2,
\e
where $ \epsilon= E/m $ is the specific energy the particle and $ j =J/m$ is its specific angular momentum.
Equation (\ref{KineticE}) is that of a one dimensional effective motion with specific energy of the one dimensional motion given by $(\epsilon^2-1)/2 $ and effective potential:

\b
V_{eff}(r) = -\frac{1}{r}M -\frac{1}{2r^2}\big[a^2(\epsilon^2-1) -j^2\big]- \frac{1}{r^3}M(a\epsilon-j)^2.
\e
To determine the boundaries of the classically forbidden region, which is characterised by negative kinetic energy, one needs to locate the roots of the equation $ R(r)= r^4\dot{r^2} = 0 $. It is instructive to consider this equation from different perspectives.
Firstly consider Carter's  equation (\ref{Carter2}) in the equatorial plane:
\b
\label{R(r)1}
R(r)=[e(r^2+a^2)-ja]^2-\Delta[r^2+(j-ae)^2].
\e
In the Kerr black hole, the time-like Killing vector $ ( \partial_t )^\mu $ becomes null at the static limit:

\b
- g_{tt} = 1-\frac{2Mr}{\rho^2},
\e
with boundary given by:

\b
r_s  = M + \sqrt{M^2- a^2 cos^2\theta }.
\e
The region between the event horizon $ r_+ $ and the static limit $ r_s$ is called the ergosphere. In this region all particles must rotate with the black hole due to frame dragging. In this region, negative energy states are possible aswell, because at the static limit $ \partial/ \partial t$ becomes space-like and can be positive or negative. Hence, the conserved energy $ E= (1/m)\partial L/\partial \dot{t} $ can be positive or negative.

In the equatorial plane, the static limit $ r_s$ is $2M$.
The location of the Cauchy horizon $ r_- $ and the event horizon $ r_+ $ are: $ r_\pm = M \pm \sqrt{M^2 - a^2 } $ respectively.
 From  (\ref{R(r)1}), it is evident that there can be no region of negative energy  between the horizons: the term\,\, $ \Delta = r^2 -2Mr + a^2 $\,\, is always negative between the horizons and hence $R(r)$ is strictly positive. A forbidden region, therefore, cannot exist between the horizons --- forbidden regions, if they exist, lie to the right of $ r_+$ or to the left of  $ r_-$.

If $\epsilon<1$, the particle cannot escape and the trajectory is bound. If $\epsilon=1$ the particle is marginally bound i.e. falling towards the hole from a state of rest at infinity and if $\epsilon >1$, the particle is unbound.
In this work, unbound particles  $\epsilon > 1 $  are considered only. Such particles with certain values of $j$ and $\epsilon$ undergo scattering by the negative kinetic energy region outside of the black hole (or tunnel across it).

Consider $R(r)$ as a cubic in $r$, i.e.:
\b
\label{KineticR}
R(r) =A r^3+Mr^2+\bigg(a^2 A -\frac{j^2}{2}\bigg)r +M(a\epsilon-j)^2,
\e
where $A= (\epsilon^2-1)/2 >0$.

   Firstly, if $a^2A > j^2/2 $, the coefficient of the $r$ term in equation (\ref{KineticR}) is positive. All other terms in the equation are positive as well ($+,+,+,+$) and so, there are no sign changes. By Descartes' rule of signs, there can only be two situations, either one negative root and two complex roots, or three negative roots. Vieta's relations result in:
   $ r_1 + r_2 +r_3 = -M/A $, \,\,  $ r_1 r_2 + r_2 r_3 + r_1 r_3 = \big( a^2 A -j^2/2 \big)/A  $,   $ r_1 r_2 r_3 = -M\big( j-a\epsilon \big)^2/A $. In this scenario, the centre is a boundary of a forbidden region if $ j= a\epsilon $.

Secondly, when $a^2A < j^2/2 $, the coefficient of the linear term in equation (\ref{KineticR}) is negative  with all others positive and one of the terms in the equation being negative ($+,+,-,+$) leads to two sign changes. There can be three situations: one negative root and two positive roots; one negative root and two complex roots; or three negative roots. The last situation cannot arise due to Vieta's restriction: $ r_1 r_2 + r_2 r_3 + r_1 r_3 = \big( a^2 A -j^2/2 \big)/A  $ --- for three negative roots the left hand side is positive while the right hand side is negative.

  Consider next the kinetic energy equation (\ref{R(r)1}) as a quadratic in $\epsilon $ and $j$ separately, both above the static limit $(r> 2M)$ and then below it  $(r_+< r< 2M)$.
Any $r$, except $r_-<r< r_+$, can lie in a forbidden region depending on the energy and angular momentum of the particle.

  Firstly, for $r>2M$, as a quadratic in $j$, the kinetic energy equation:

\b
\label{Rj}
R(j)=  \big(2M - r  \big)j^2 -4\epsilon M aj- \Delta r + \epsilon^2 (r^3 + a^2 r + 2a^2 M),
\e

  leads to a  negative kinetic energy region for values of $j$ between the roots:

\b
\label{Jr}
j_\pm = \frac{2a\epsilon M}{(2M-r )} \bigg[1 \pm \sqrt{1-\frac{\big(\epsilon^2r^3-\Delta r+a^2 \epsilon^2 (r+2M) \big)(2M-r)}{4a^2 \epsilon^2M^2}  } \bigg ].
\e

 The location of these roots is determined by considering the sign changes in (\ref{Rj}). The quadratic  term is always negative, the linear term  can be positive or negative, depending on the sign of  $ aj$ ($ \epsilon >0$ always) and the constant term can also be positive or negative depending on $r$. If $aj$  is positive (co-rotating probe and centre) then the signs are $-,-,-,$ or $-,-,+,$ and there is either one positive and one negative root, or two complex roots. If $aj$ is negative (counter-rotating probe and centre) then all possibilities may arise.

Furthermore, for real roots to exist, the kinetic energy of the particle must be above the threshold energy:

\b
\label{Er1}
\epsilon  > \epsilon_T= \sqrt{ \frac{r-2M}{r}}.
\e
 Only then the discriminant of (\ref{Rj}) is positive.(Fig. 1a) For  particles with energy greater than $\epsilon_T$ and angular momentum in the range $( j_-,j_+)$, there is no forbidden region. For particles with $\epsilon > \epsilon_T$ and  angular momentum outside the interval $( j_-$, $ j_+)$  or with $\epsilon < \epsilon_T $ and any $j$, only circular motion is possible, since the forbidden region extends for all $r>2M$.

Next, consider the kinetic energy equation (\ref{R(r)1}) as a  quadratic in $\epsilon $ for $r>2M$:

\b
\label{Re}
R(\epsilon)=   (r^3 + a^2 r + 2a^2 M) \epsilon^2  -4ajM\epsilon  -[\Delta r +(r - 2M)j^2] .
\e

A negative kinetic energy region exists for energies between the roots:

\b
\label{Er}
\epsilon_\pm = \frac{2 a j M}{r^3 + a^2( r + 2M) }\bigg[ 1\pm \sqrt{1- \frac{\big(  -\Delta r-j^2\big(r+2M\big)\big)\big[r^3+ a^2( r +2M)  \big]}{4a^2j^2 M}  }  \bigg].
\e

In this equation, the quadratic term is always positive as is the constant term ($r> 2M$). The linear term is positive or negative depending on the sign of $aj$. The signs of the terms are therefore either $+,+,-,$ or $+,-,-,$ so there is only one sign change in each case. As the discriminant is positive for all values of $j$, the only possible roots are one positive and one negative (Fig. 2a).

Particles with kinetic energy in the range ($ 0, \epsilon_+$) \footnote{Equation (\ref{Re}) is generic: valid for all $\epsilon >0$ (not for unbound particles only). } and any angular momentum experience a negative kinetic energy region [$R(\epsilon)<0$] and their radial motion is restricted. For $\epsilon > \epsilon_+$ and any $j$, then $R(\epsilon) >0$, and there is no forbidden region.

Secondly, below the static limit $(r<2M)$, consider the sign changes of the  kinetic energy equation $(\ref{Rj})$. In this case the quadratic term is always positive, while the signs of the other terms can be positive or negative. For $aj>0$, the possible signs of the terms are $+,-,+,$ and $+,-,-$. There are three possibilities: two positive roots, one positive root and one negative root, or two negative roots. When $aj<0$ the possible signs of the terms are $+,+,+,$ and $+,+,-$. In this case there is either one positive root and one negative root or two negative roots. The expression for the roots is the same as the case where $r >2M$ $(\ref{Jr})$. The discriminant is positive for all $\epsilon$ in this case. So, in the ergosphere, for any $\epsilon $ and for angular momentum outside the range $(j_-, j_+)$, $R(j) >0$ and a forbidden region does not  exists(Fig. 1b)

Now consider again the kinetic equation as a quadratic in $\epsilon$ $(\ref{Re})$. The signs of the coefficients of the quadratic  and linear terms are the same as in the previous case $(r>2M)$. However the constant term can now be positive or negative depending on $r$. For positive values of $aj$ the signs are $+,-,+, $ and $+,-,- $. For negative values of $aj$ the signs are $+,+,+, $ and $+,+,- $.  The possibilities are two positive roots or one positive and one negative root ($j>0$) or one positive and one negative or two negative root (no complex roots are possible in this case since the discriminant is positive) ($aj<0$). Therefore, in the ergosphere a negative kinetic region $R(\epsilon)< 0$ exists for particles with energies in the range  $(\epsilon_-, \epsilon_+)$ and any angular momentum.(Fig. 2b)

Particles with kinetic energy and angular momentum within the ranges outlined above experience a negative kinetic energy region  and their radial motion is restricted. Outside these ranges of energies  angular momentum Hawking radiation particles will escape to infinity and incoming particles will fall into the black hole(Fig. 3).

\begin{figure}[h]
\centerline{
\includegraphics[width=0.45\textwidth, height=0.45\textwidth]
{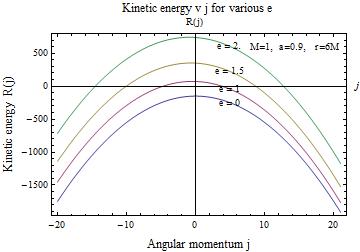}
\hspace{1cm}
\includegraphics[width=0.45\textwidth, height=0.45\textwidth]
{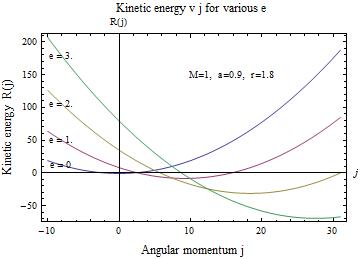} }
\caption{\label{figs1}(a) Kinetic energy v $j$ $(r>2M)$.
\hspace{1cm} (b) Kinetic energy v $j$ $(r<2M)$.}

\end{figure}

\begin{figure}[h]
\centerline{
\includegraphics[width=0.45\textwidth, height=0.45\textwidth]
{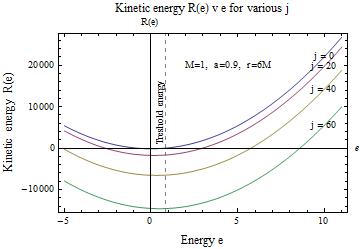}
\hspace{1cm}
\includegraphics[width=0.45\textwidth, height=0.45\textwidth]
{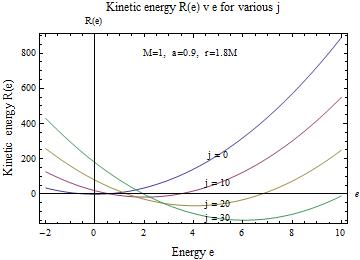} }
\caption{\label{figs1} (a) Kinetic energy v energy  $( r> 2M) $.
\hspace{1cm} (b) Kinetic energy v energy $(r < 2M)$.}

\end{figure}

\begin{figure}[h]
\centerline{
\includegraphics[width=0.45\textwidth, height=0.45\textwidth]
{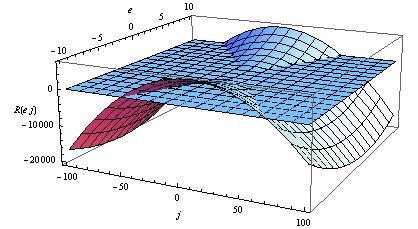}
\hspace{1cm}
\includegraphics[width=0.45\textwidth, height=0.45\textwidth]
{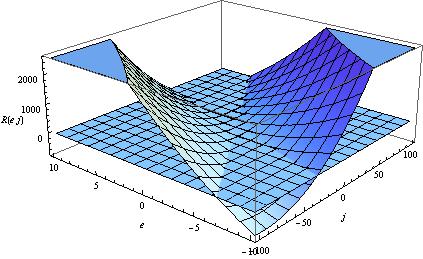} }
\caption{\label{figs1} (a) Kinetic energy v $\epsilon$ and  $j (r>2M)$.
\hspace{1cm} (b) Kinetic energy v $\epsilon$  and $j  (r<2M)$.}

\end{figure}

  To summarise, for values of $ j $ and $ \epsilon $, a forbidden region exists to the right of the event horizon $ r_+ $. For a forbidden region to exist above the event horizon $(r_+)$ and within the ergosphere (below the static limit $2M$) the particle's kinetic energy should be in the  range ($ \epsilon_-$, $ \epsilon_+$) for any angular momentum  or its angular momentum should be in the range $ (j_-, j_+) $ for any energy. For a particular value of energy and angular momentum a double root appears just above $r_+$. As $j$ increases further two real roots $r_2$ and $r_3 $  appear, forming the boundaries of the forbidden region.

   The first of these, above the event horizon $r_+$, occurs for particular values of $j$ and $\epsilon$ at $r_2=r_+ + \delta_+$.  From equation (\ref{R(r)1}), the condition under which the root approaches the event horizon ($\delta_+ \rightarrow 0 $) can be determined by requiring:

\b
e(r^2+a^2)-ja \rightarrow 0,
\e
    or, $ j \rightarrow 2M \epsilon r_+/a$, as a root approaches the horizon from above and $ j \rightarrow 2M  \epsilon r_-/a$, as a root approaches the Cauchy horizon from below ($\delta_- \rightarrow 0$).

    The other boundary of the forbidden region is given by the root $r_3$. As the angular momentum of the particle increases or, as the energy decreases, this boundary extends above the static limit $r>2M$. Here, a forbidden region exists for a particle with angular momentum in the interval $ (-\infty, j_-]\cup [j_+, \infty) $ and for kinetic energies above a threshold energy ($ \epsilon_T $). For particles with energy less that $\epsilon_T $  and any angular momentum no radial motion is possible since any $ r< \infty$ could be in a forbidden region for certain values of $\epsilon$ and $j$.

  The root $r_3$ can be determined in terms of $r_2$ by eliminating $r_1$ using Vieta's relations: $ r_1 + r_2 +r_3 = -M/A $,   $ r_1 r_2 + r_2 r_3 + r_1 r_3 = \big( a^2 A -j^2/2 \big)/A  $,   $ r_1 r_2 r_3 = -M\big( j-a\epsilon \big)^2/A $:

\b
r_3^2 + r_3\bigg(\frac{M}{A} +r_2 \bigg)-\frac{M}{r_2 A}\bigg( J-a\epsilon \bigg)^2 = 0.
\e
Only the positive root $r_3$ of this equation is of interest. The lower boundary of the forbidden region above the event horizon is at $r_2=r_+ +\delta_+ $. Thus the upper boundary is at:

\b
  r_3 & = & \frac{1}{2}\bigg(\frac{2M + (r_+  +\delta_+)(\epsilon^2 -1)}{\epsilon^2 -1}  \bigg)  \\
  & & \times  \Bigg[-1 + \sqrt{1+ \frac{8M(j-a\epsilon)^2(\epsilon^2-1)}{(r_+ +\delta_+)\big( 2M +(r_+ +\delta_+)(\epsilon^2-1) \big)^2}}  \Bigg].  \nonumber
\e

Similarly, below the Cauchy horizon the upper boundary of the forbidden region extends to $ r_2 = r_- - \delta_-$. and the lower boundary is at $r_3$:
\b
r_3 & = & \frac{1}{2}\bigg(\frac{2M + (r_-  -\delta_-)(\epsilon^2 -1)}{\epsilon^2 -1}  \bigg)  \\
& & \times \Bigg[-1 + \sqrt{1+ \frac{8M(j-a\epsilon)^2(\epsilon^2-1)}{(r_- -\delta_-)\big( 2M +(r_- -\delta_-)(\epsilon^2-1) \big)^2}}  \Bigg]. \nonumber
\e

By expanding the kinetic energy equation (\ref{KineticE}) in a power series near the Cauchy and near the event horizon, the values of $\delta_\pm$ can be  approximated by:

\b
\delta_\pm = \frac{r_\pm\big[\big(\epsilon M r_\pm - aj^2 \big)^2 + \epsilon^2 M r_\pm ^2  \big]}{\big(\big(\epsilon^2 -1\big)r_\pm^3-j^2r_\pm  \big)\big( r_+-M \big) \pm \big( \epsilon M r_\pm + a j\big)\big(2 \epsilon M r_\pm + 4a j \big)}.
\e

Two forbidden regions may exist depending on the values of $j $ and $ \epsilon$. The region outside the event horizon screens the Hawking radiation and shields the Penrose process.  Classically this is a barrier to particles escaping the black hole or entering the ergosphere and contributing to the decay of the black hole. However, particles may tunnel across quantum mechanically  and offset these effect.

\section{ Hawking radiation. Screening reduction }	
Particle pairs are produced due to vacuum fluctuations inside the event horizon  at $r=2M-\delta $ (where $\delta$ is small). One of them, the ingoing particle, moves towards the centre of the black hole. The other, the outgoing particle, tunnels across the horizon and materialises outside at $r=2(M-\omega)+\delta$\cite{Parikh1}.

The imaginary part of the action for an outgoing positive energy particle is:

\b
Im  S = Im\int_{r_{in}}^{r_{out}} p_rdr,
\e
where $ r_{in} $ and $ r_{out}$ are given by:

\b
r_{in} = M + \sqrt{M^2-a^2},
\e
\b
r_{out} = M-\omega + \sqrt{(M-\omega)^2-a^2},
\e
and $\omega$ is the energy of the emitted particle.

For the Kerr black hole:

\b
Im S=\frac{\pi(r_+^2 +a^2)}{r_+-M}\omega,
\e
and the Hawking tunnelling rate is:
\b
\Gamma = e^{-2Im S}.
\e
The average energy of the particles emitted from the surface of the black hole is:

\b
E = \frac{3}{2}k T.
\e
The temperature given by:

\b
T = \frac{\kappa}{2\pi},
\e
where $\kappa$ is the surface gravity of the black hole \cite{Jacobson}:
\b
\kappa = 4\pi \frac{\mu}{A}.
\e
Here $A$ is the horizon area of the Kerr black hole \cite{Bardeen}:

\b
A= 4\pi[2M(M+\mu)],
\e
and

\b
\mu =\sqrt{M^2-\frac{a^2}{M^2}}.
\e
The average energy of the particles emitted via Hawking radiation from the Kerr black hole is therefore:

\b
\label{HE}
   E =  \frac{3}{8 \pi}\frac{\sqrt{M^2-\frac{a^2}{M^2}}}{M\big(M +\sqrt{M^2-\frac{a^2}{M^2}}\big)}
\e
Depending on the spin and the mass of the black hole, the angular momentum and energy of the emitted particles, screening of such particles may take place.

\begin{figure}[h]
\centerline{
\includegraphics[width=0.45\textwidth, height=0.45\textwidth]
{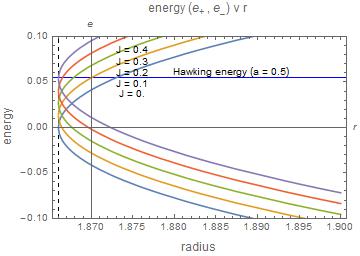}
\hspace{1cm}
\includegraphics[width=0.45\textwidth, height=0.45\textwidth]
{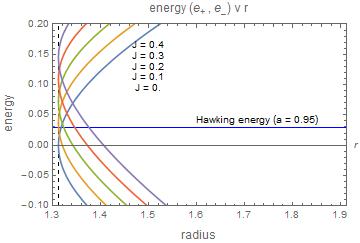} }
\caption{\label{figs1}(a) Energy  $v$ $r$ $(a=0.5)$.
\hspace{1.5cm} (b) Energy  v $r$ $(a= 0.95)$. \hspace{2cm} Here the effect of the angular momentum on Hawking screening is demonstrated. Hawking particles with energy $E = $0.55 and $j >$ 0.4 are screened while lower angular modes at this energy will escape the event horizon (Fig 4a). For higher spin ($a=$ 0.95) the energy of the Hawking particles is $E =$ 0.02 and screening of all but ($j$ = 0) angular momentum modes takes place (Fig.4b). }

\end{figure}

\begin{figure}[h]
\centerline{
\includegraphics[width=0.45\textwidth, height=0.45\textwidth]
{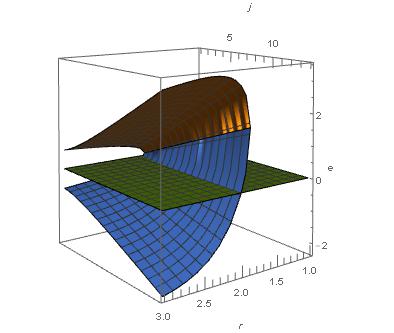}
\hspace{1cm}
\includegraphics[width=0.45\textwidth, height=0.45\textwidth]
{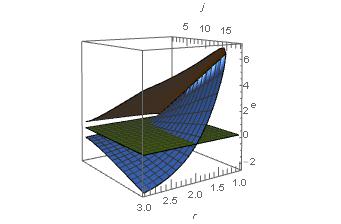} }
\caption{\label{figs1}(a) Energy  v $r$ and $j$ $(a=0.5)$.
\hspace{1.5cm} (b) Energy  v $r$ and $j$  $(a= 0.95)$. \hspace{2cm} The screening of Hawking particles with energy ($E$ = 0.055) from a Kerr black hole ( spin $a$ = 0.5) takes place in the green region between the two surfaces (Fig.5a). The effect is greatly enhanced  for higher spin value ($a$ = 0.95) and  lower energy $E $ = 0.02 (Fig. 5b). }

\end{figure}

 A negative kinetic energy region exists outside the horizon for particles with energies in the range ($\epsilon_-, \epsilon_+$) (\ref{Er}) and angular momenta in the range ($j_-,j_+$)  (\ref{Jr}). Emitted particles with such energies and angular momenta are reflected back into the black hole at the lower boundary $r_2$ of the forbidden  region and screening takes place. This is evident for black holes with spin ($a$ = 0.5). Particles with energy $E$ = 0.055 and angular momentum less than $j=$ 0.4 will escape the event horizon, while particles with greater angular momentum will be reflected back into the horizon (Fig 4, 5). The energy for various angular momenta are given in the table (Table 1).

 \begin{table}
  \begin{center}
  \caption{ Hawking radiation screening in the Kerr black hole }
\begin {tabular}{|*{7}{l}|}
\hline
Angular &\multicolumn{2}{c}{ $j$ = 0}  & \multicolumn{2}{c}{ $j$ = 0.2 }  &\multicolumn{2}{c|}{ $j$ = 0.4}  \\
momentum   & & & & & & \\
\hline
radius & $\epsilon_+ $& $ \epsilon_-$ &  $\epsilon_+$ &$ \epsilon_-$ & $\epsilon_+$&$\epsilon_-$\\
\hline\hline
1.87 & 0.0414 & -0.0414 & 0.0683 &-0.0150 & 0.0955 &0.1810 \\

1.88 & 0.0776 & -0.0776 & 0.1042 & -0.0517 & 0.1316 & -0.0266\\

1.9& 0.1206 &-0.1206  & 0.1467 & -0.0956 & 0.1740 & -0.0718\\

\hline
1.92 & 0.1515 & -0.1515 & 0.1771 & -0.1274  & 0.2040 & -0.1047\\
1.94 & 0.1768 & -0.1768 & 0.2018 & -0.1535 &0.2284 & -0.1318\\
1.96 & 0.1896 & -0.1986 & 0.2230 & -0.1761 & 0.2492 & -0.1553\\
\hline
1.98 &0.2181 & -0.2181 & 0.2149& -0.1962  & 0.2676 & -0.1763\\
2.0 & 0.2357 & -0.2357 & 0.2589 & -0.2145 & 0.2843 & -0.1954 \\

\hline
\hline
\end{tabular}
\end{center}

Table 2: Screening of Hawking radiation for the Kerr black hole with spin ($a = 0.5$), event horizon at 1.866 and $E $ = 0.055 (Hawking energy). All but the very lowest angular momentum modes are screened.
\end{table}

  Some particles, however, may tunnel across this potential barrier and escape the black hole. This reduction in screening depends on the width $w=r_3-r_2 $ of the forbidden region. As the angular momentum grows or as the energy decreases, this region widens. The effect alters the spectrum of the Hawking radiation and dampens higher angular momentum spectral modes and low energy modes. Hawking radiation is therefore dominated by low angular momentum modes.

\section{ Penrose process. Shielding }

According to Penrose \cite{Penrose}, it is possible to extract from a rotating black hole. Consider a particle falling from infinity into a Kerr black hole and splitting into two fragments.
The Penrose process occurs when a particle with negative energy is absorbed by the black hole. As a result the hole's mass and angular momentum decrease. If the process continues, eventually the Kerr black hole would turn into a Schwarzschild black hole. The process depends on negative energy geodesics in the ergosphere.
Consider a particle $P_0$ with energy $ E_0$ entering the ergosphere and splitting into two with the first particle $P_1$ having energy $ E_1$ and the second particle $ P_2$ having energy $ E_2$. The two particles will fly off in opposite directions, $ P_1$ going in the direction of rotation of the black hole and $P_2$ going in the opposing direction. $P_1$ goes into a negative energy orbit and falls into the black hole and $P_2$ emerges form the ergosphere with increased energy $ E_2$.

Effectively, $ P_2$ gains energy from the ergosphere:  $ E_2 > E_0 $ --- a transfer of energy from the hole to $P_2$.
 Chandrasekhar determined the efficiency of the process (the ratio of the maximum energy out to the energy in) to be 20.7.  \cite{Chandrasekhar} Increased energy extraction efficiency has been demonstrated in recent years by considering modifications to the process. Banados, Silk and West (BSW)\cite{Banados} considered colliding particles near the event horizon of an extremal black hole resulting in higher energy extraction efficiency due to the high centre-of-mass energy of the collision. They proposed that Kerr black holes can act as natural particle accelerators.  More recently, Schnittman \cite{Schnittman} has demonstrated a new limit of energy extraction efficiency for Compton scattering between a photon and a massive particle. In this mechanism the efficiency can reach 1400\%.

The effectiveness of the Penrose process is reduced for particles with energies ($\epsilon_-<\epsilon <\epsilon_T$) (\ref{Er}) and angular momenta within the range ($j_-,j_+$) (\ref{Jr}) for $r<2M$. For particles with energies ($\epsilon >\epsilon_T$) and angular momenta outside the range ($j_-,j_+$) the entire ergosphere is in the forbidden region ($r>2M$) and there is no Penrose process. Three scenarios are relevant.

   Firstly, for particles with energies  ($\epsilon_-< \epsilon <\epsilon_T$) (\ref{Er}) and for angular momenta within the range ($j_-,j_+$) (\ref{Jr}), the boundary of the forbidden region is below the static limit $ r_3< r_s $ and partial shielding occurs. Particles may enter the ergosphere and energy extraction or particle acceleration can take place. As the angular momentum increases or the energy decreases, the boundary $r_3$ approaches the static limit $r_s$ and the area available for energy extraction decreases.

   Secondly however, tunneling can take place across the potential barrier and the narrower the forbidden region, the more likely the tunnelling. Some of the particles tunneling across continue on through the event horizon $r_+$ and will not take part in the process. Other particles, however, may still undergo the Penrose process between the event horizon $r_+$ and the forbidden region $r_2<r<r_3$. The negative energy particle falls into the event horizon and slows down the black hole. The particle gaining energy from the ergosphere may tunnel back across the potential barrier, exit the black hole and contribute to the spin-down process.  The exiting particle may not tunnel across the potential barrier but may be reflected back at the boundary $r_2$ into the event horizon and thus spin the hole up.

Finally, for particles with ($\epsilon <\epsilon_T$) or for particles with  $j$ outside  $(j_-,j_+)$ when ($\epsilon_>\epsilon_T$) the negative kinetic energy region extends beyond the static limit ($r_3>r_s$) and the black hole is completely shielded. It becomes "invisible" to these particles and the Penrose process cannot take place. However, particles can still tunnel across the potential barrier and  enter the region above the event horizon and below the forbidden region: $r_+ <r<r_2$. In a repeat of the above scenario a greatly reduced Penrose process takes place.

The shielding effect of a maximal rotating kerr black hole on a freely falling particle at the staic limit is demonstrated in figure 6a. The shielding effect beyond the static limit is illustrated in terms of the impact parameter (Fig.6b) and in terms of negative kinetic energy regions for various values of angular momentum and energy (Table 2).
\begin{figure}[h]
\centerline{
\includegraphics[width=0.45\textwidth, height=0.45\textwidth]
{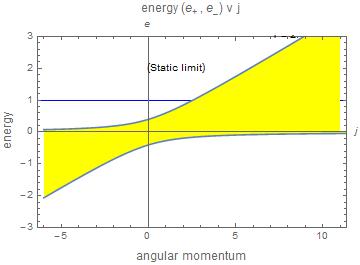}
\hspace{1cm}
\includegraphics[width=0.45\textwidth, height=0.45\textwidth]
{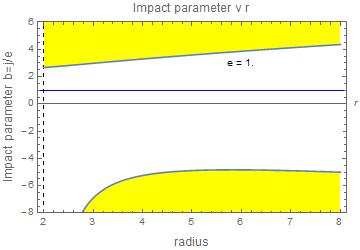} }
\caption{\label{figs1}(a) Energy  v $r$ $(a=1.0)$.
\hspace{2.0cm} (b) Energy  v $r$ $(a= 1.0)$.   \hspace{2cm}  In (Fig. 6a) particles with energies and angular momenta in the the yellow region are shielded at the static limit for a Kerr Black hole with maximal spin $a$ =1.  Beyond the static limit particles in the yellow region will be shielded from the black hole for values of $j $ outside the interval($j_-$, $j_+$).}

\end{figure}

 \begin{table}
  \begin{center}

  \caption{Penrose process shielding in the Kerr black hole}
\begin {tabular}{|*{7}{l}|}
\hline
Energy & \multicolumn{2}{c}{ $\epsilon$ = 1.0} & \multicolumn{2}{c}{ $\epsilon$ = 1.2 }  & \multicolumn{2}{c|}{ $\epsilon$ = 1.4}  \\
radius & $ j_+ $& $ j_-$ &  $j_+$ &$ j_-$ & $j_+$&$j_-$\\
\hline\hline
2.5 & -10.5689 & 2.7891 & -12.9818 & 3.5783 & -15.3499 & 4.3739 \\

3.0 & -6.8437 & 2.9147 & -8.6325 & 3.9285 & -10.3585 & 4.8705 \\

3.5 & -5.7022 & 3.0889 & -7.4159 & 4.2799 & -9.0252 & 5.3665 \\
4.0 & -5.2133 & 3.2533 & -6.9804 & 4.6284 & -8.6058 & 5.8618\\
\hline
4.5 & -4.9772 & 3.4092 & -6.8560 & 4.9744  & -8.5516 & 6.3564\\
5.0 & -4.864 & 3.5578 & -6.8865 & 3.3185  & -8.6796 &  6.8503\\
5.5 & -4.8200 & 3.7000 & -7.0050 & 5.6610 & -8.9116 & 7.3436 \\
6.0 & -4.8167 & 3.8367 & -7.1782 & 6.0022 & -9.2086 & 7.8366\\
\hline
6.5 & -4.8394 & 3.9683 & -7.3876 & 6.3423  & -9.5487  & 8.3216 \\
7.0 & -4.8792 & 4.0955 & -7.6223 & 6.6813 & -9.9190 & 8.8214 \\
7.5 & -4.9313 & 4.2186 & -7.8752 & 7.0199  & -10.3112 & 9.3134\\
8.0  & -4.9914 & 4.3381 &  -8.1417 & 7.3577 & -10.7199 &  9.8051\\
\hline
\hline
\end{tabular}
\end{center}

Table 2: Regions of negative kinetic energy ($j_+ > j > j_-$) outside the ergosphere of the Kerr black hole ( $a = 0.98$) shield particles with various energies and  angular momenta.
\end{table}

\section{ Tunneling }

Now consider the one-dimensional Schr\"{o}dinger equation for a particle with energy $E$ and moving in potential $V$ \cite{Gamov}:
\b
\label{Scho}
 \frac{d^2\psi }{dr^2} =-\frac{p^2(r)}{h^2}\psi(r),
\e
where $ p^2(r) = 2m[E-V(r)] $ is the square of the classical momentum $ p(r)$ of the particle. For particles with $ E < V(r)$, quantum mechanical-tunneling can take place across the classically forbidden region bounded by the turning radii $ r_2, r_3 $. The wave function, determined  by the  WKB approximation method, is\cite{Gamov}:

\b
\psi(r) = \frac{D}{\sqrt{|p(r)|}} e^{\pm\frac{1}{\hbar}\int_a^b |p(r)|dr},
\e
where D is a constant. The amplitude of the incident wave is attenuated on transmission across the potential barrier --- decreased by the factor $ e^{2\gamma} $, where:

\b
\label{gamma}
\gamma =\frac{1}{\hbar}\int_{r_2}^{r_3}|p(r)|dr.
\e
The tunneling probability is proportional to the Gamow factor $ e^{-2\gamma}  $.

In the presented setup, the forbidden region corresponds to the area of negative specific kinetic energy, where:

\b
p^2(r) = 2[V_{eff}(r) - \frac{\epsilon^2 -1}{2} ]= -\dot{r}^2.
\e
Thus:

\b
\gamma =\frac{\sqrt{2m}}{\hbar}\int_{r_2}^{r_3}\sqrt{[-M(a\epsilon-j)^2\frac{1}{r^3} -(a^2(\epsilon^2-1) -j^2 )\frac{1}{2r^2}-M\frac{1}{r}-\frac{\epsilon^2-1}{2} ]}dr. \nonumber
\e
\newline

In order to evaluate this integral the cubic expression under the square root can be approximated by a quadratic in the region of interest only. That is, a parabolic function $ W(r)$ whose roots $R_{\pm}$ and local maximum coincide with with those of the cubic polynomial, Hence:

\b
V_{eff}(r) - \frac{\epsilon^2 -1}{2} \simeq W(r) &=& T\bigg(\frac{1}{r}-\frac{1}{R_-} \bigg)\bigg(\frac{1}{r}-\frac{1}{R_+}\bigg) \\
     &=& \frac{T}{r^2R_-R_+}(r-R_- )(r-R_+). \nonumber
\e
The approximating parabola will then go through the point $ ( s,R(s)) $, where:
\b
s =  \frac{r_3 - r_2 }{2}.
\e
This condition is satisfied when T is given by:
\b
T = s^2R_-R_+\frac{V_{eff}(s)-\frac{\epsilon^2-1}{2}}{(s-R_-)(s-R_+)},
\e
where $r_2 = R_-$ and $r_3= R_+$.
Here $ R_- $ and $ R_+$ are  the roots of the approximating parabola and also upper and lower bounds of the forbidden region.

Within this approximation, the $\gamma$ factor becomes:

\b
\gamma &=& \frac{\sqrt{2m}}{\hbar}\Big[ s^2r_2r_3\frac{V_{eff}(s)-\frac{\epsilon^2-1}{2}}{(s-r_2)(s-r_3)} \Big]^\frac{1}{2}\int_{r_2}^{r_3}\sqrt{\frac{(r-r_2)(r-r_3)}{r^2}} dr \nonumber \\
 &=& \Big[ \frac{m\pi^2}{4h^2}r_2r_3\frac{V_{eff}(s)-\frac{\epsilon^2-1}{2}}{(r_3-3r_2)} \Big]^\frac{1}{2}(\sqrt{r_3}-\sqrt{r_2})^2 \nonumber \\
 &=& \Big[ \frac{m\pi^2}{4h^2}\frac{r_2(r_2+w)}{(w-2r_2)} \Big]^\frac{1}{2} \frac{w^2}{(\sqrt{r_2+w}+\sqrt{r_2})^2} \nonumber \\
 &\times& \Big [-8M(a\epsilon-j)^2\frac{1}{w^3} -(2a^2(\epsilon^2-1) -j^2 )\frac{1}{w^2}-M\frac{1}{w}-\frac{\epsilon^2-1}{2} \Big ]^ \frac{1}{2},
\e
where $w$ is the width of the potential barrier.

The Gamow factor  $ e^{-2\gamma} $ is directly related to the width $w$ of the potential barrier and also the angular momentum and energy of the tunnelling particle. $\gamma$ increases with the square of the barrier width. In the limit $ w \rightarrow \infty$, the tunnelling probability (which is proportional to $ e^{-2\gamma} $) diminishes exponentially. As the width of the potential barrier decreases:  $ w\rightarrow 0$, $\gamma \rightarrow 0$ and the likelihood of tunnelling is increased.

\section{ Conclusions}

Restrictions to the Hawking and Penrose decay processes in the Kerr black hole occur due to screening of particles with a range of energies and a range angular momenta. The effect takes place at the boundaries of a negative kinetic energy region which may exist outside the event horizon and affects both the Hawking radiation and the Penrose process. The locations of these boundaries are determined and found to depend on the angular momentum and energy of the particles. Tunneling across this region or potential barrier may take place, however, resulting in reducing these effects.

The screening of Hawking radiation particles results in the attenuation of parts of the energy spectrum and of the angular momentum spectrum. In a competing process, tunnelling offsets the screening and allows a fraction of these particles to be transmitted  through the potential barrier. This tunneling is reduced when the angular momentum of the particles increases or the energy decreases and the width of the forbidden region widens. As a result, the Hawking radiation signature is predominantly made up of low angular momentum modes or high energy modes.

The Penrose process is reduced for particles with low energies and high angular momenta due to the shielding effect of the forbidden region and hence the reduction of energy extraction. As the angular momentum of the particles increases, or their energy decreases, the black hole becomes completely shielded. Some particles can still tunnel across the barrier offsetting the shielding process.

It is intended also to extend the study to the Kerr-Newman and Kerr-de Sitter cases.
Future work will involve a more quantitative approach to these decay processes with a comparative study of black holes from the microscopic to the astronomical.

\end{document}